# Evolution of superconductivity in $LaO_{1-x}F_xBiS_2$ prepared by high pressure technique

K. Deguchi[1,2,3], Y. Mizuguchi[1,2,4], S. Demura[1,2,3], H. Hara[1,2,3], T. Watanabe[1,2,3],
S. J. Denholme[1,2], M. Fujioka[1,2], H. Okazaki[1,2], T. Ozaki[1,2], H. Takeya[1,2], T. Yamaguchi[1,2],
O. Miura[4], and Y. Takano[1,2,3]

[1]*National Institute for Materials Science, 1-2-1, Sengen, Tsukuba, 305-0047, Ibaraki, Japan*
[2]*JST-EU-JAPAN, 1-2-1, Sengen, Tsukuba, 305-0047, Ibaraki, Japan*
[3]*University of Tsukuba, 1-1-1, Tennodai, Tsukuba, 305-8577, Ibaraki, Japan*
[4]*Tokyo Metropolitan University, 1-1, Minami-osawa, Hachioji, 192-0397, Tokyo, Japan*

Novel $BiS_2$-based superconductors $LaO_{1-x}F_xBiS_2$ prepared by the high pressure synthesis technique were systematically studied. It was found that the high pressure annealing strongly the lattice as compared to the $LaO_{1-x}F_xBiS_2$ samples prepared by conventional solid state reaction at ambient pressure. Bulk superconductivity was observed within a wide F-concentration range of $x = 0.2 \sim 0.7$. On the basis of those results, we have established a phase diagram of $LaO_{1-x}F_xBiS_2$.

## 1. Introduction

Recently, several $BiS_2$-based superconductors, commonly having the Bi-S square lattice planes, have been discovered.[1-15] Due to the layered crystal structure and some exotic physical properties similar to cuprate[16-19] and Fe-based superconductors,[20-31] the $BiS_2$-based compounds are expected to provide us with the next stage to explore new superconductors and discuss the exotic superconductivity mechanisms. The $Bi_4O_4S_3$ superconductor exhibits metallic transport behavior and show a zero-resistivity state below 4.5 K.[1] The crystal structure is composed of a stacking of the $Bi_4O_4(SO_4)$ blocking layers and the $Bi_2S_4$ superconducting layers (two $BiS_2$ layers). Thus, the parent phase is $Bi_6O_8S_5$ and it is expected to be insulator on the basis of the band calculations. The $Bi_4O_4S_3$ phase has partial defects at the $SO_4$ site, which provide electron carriers into the $BiS_2$ superconducting layers. Another $BiS_2$-based system is $ReO_{1-x}F_xBiS_2$ (Re = Rare earth). So far, $LaO_{1-x}F_xBiS_2$, $CeO_{1-x}F_xBiS_2$, $PrO_{1-x}F_xBiS_2$ and $NdO_{1-x}F_xBiS_2$ were found to be superconducting with transition temperatures ($T_c$) of 10.6[2], 3.0[14], 5.5[15], 5.6 K[11], respectively. In both systems, optimal superconducting properties are obtained near the boundary between insulating and superconducting states. In fact, the electronic-specific-heat coefficient of the $Bi_4O_4S_3$ superconducting sample was found to be very small.[32] This respect resembles to the layered nitride family.[33,34] By theoretical studies, possible paring mechanisms relating to charge-density-wave instability and nature of strong coupling were predicted.[9] Although the superconductivity mechanisms of the $BiS_2$-based family are unclear, we can expect a higher $T_c$ in this system, because of some exotic physical and structural properties. In fact, an enhancement of $T_c$ under high pressure was observed in $LaO_{1-x}F_xBiS_2$ system.[35] Therefore, systematic studies of both structural and superconducting properties are important. In this article we report systematic studies on $LaO_{1-x}F_xBiS_2$ superconductors prepared using a high pressure synthesis technique.

*DEGUCHI.Keita@nims.go.jp

## 2. Experimental details

The polycrystalline samples of $LaO_{1-x}F_xBiS_2$ were prepared by two-step process of the solid state reaction and the high pressure annealing using a Cubic-Anvil-type high pressure synthesis machine with a 180 ton press. The starting materials of $Bi_2O_3$ (98 % powder), $BiF3$ (99.9 % powder), $La_2S_3$ (99.9 % powder), Bi (99.9 % grains) were used in this study. The $Bi_2S_3$ powder was prepared using Bi grains and S (99.9999 %) grains. The starting powders with a nominal ratio of $LaO_{1-x}F_xBiS_2$ with $0 \leq x \leq 0.7$ were well-mixed and pressed into pellets. The pellets were sealed into an evacuated quartz tube and heated at 700 °C for 10 h. The obtained pellets were ground and annealed at 600 °C for 1 h under a hydrostatic pressure of 2 GPa. The obtained samples were characterized by x-ray diffraction with Cu-K$\alpha$ radiation using the 2$\theta$-$\theta$ method. Lattice parameters were calculated using the least-square calculations. The electrical resistivity was measured using the four-terminal method from 300 to 2 K. The magnetic susceptibility measurements were performed using a superconducting quantum interference device SQUID magnetometer from 15 to 2 K. The magnetic susceptibility measurements were performed after both zero-field-cooling (ZFC) and field-cooling (FC) with an applied field of 10 Oe. In this article, we classify synthesis methods "HP" and "AP", which stand for high-pressure-annealed and ambient-pressure annealed samples, respectively.

## 3. Results and discussion
### 3.1 Crystal structure

Figure 1 shows the powder x-ray diffraction patterns for $x = 0 \sim 0.7$ (HP). Almost all of the peaks are indexed using the space group of P4/*nmm*. For lower $x$, the pattern and peak sharpness seem to be relevant to those of AP samples. With increasing $x$, however, the peaks become broader. To compare the peak shifts, we plotted the enlarged patterns near the (102) and (004) peaks for $x = 0$ (HP), 0.2 (HP) and 0.5 (HP) with those for $x = 0$ (AP), 0.2 (AP) and 0.5 (AP) in Fig. 2. For both the AP and HP samples, clear peak shifts corresponding to lattice shrinkage with increasing F concentration. Interestingly, we note an obvious deference in between the powder patterns for $x = 0$ (AP) and 0 (HP). The (102) peak position of $x = 0$ (HP) is clearly higher than that of $x = 0.5$ (AP), while the (004) peak position seems to show a slight shift. These facts indicate that the high-pressure annealing can shrink the *ab* plane as compared to the AP synthesis. The calculated lattice constants *a*, *c* and volume (*V*) are plotted as a function of $x$ in Figs. 3(a), 3(b) and 3(c), respectively. In Fig. 3(a), it is found that the a parameters of HP samples are smaller than those of AP samples. The $x$ dependence of a parameter exhibits a dome-shaped dependence for the HP samples. In contrast, the *c* axis and lattice volume shows a continuous decrease with increasing $x$.

*3.2 Superconducting properties*

Figure 4(a) shows the temperature dependence of resistivity from 300 down to 2 K for $LaO_{1-x}F_xBiS_2$ with $x = 0 \sim 0.7$. For $x = 0$, a semiconducting-like behavior is observed and superconducting transition is not detected above 2 K. An enlargement of low temperatures below 15 K is shown in Fig. 4(b). With F doping, the semiconducting-like behavior is slightly suppressed and superconductivity appears in $x = 0.2$. With further F-doping, the semiconducting-like behavior is enhanced again. However, the $T_c$ is enhanced and exceeds 10 K (onset) at $x = 0.5$. Then, superconductivity is gradually suppressed for $x > 0.5$ and disappears at $x = 0.7$. Correspondingly to the resistivity measurements, the evolution of bulk superconductivity is also confirmed by magnetic susceptibility measurements. Figure 5(a) shows the temperature dependence of magnetic susceptibility below 12 K for $LaO_{1-x}F_xBiS_2$ with $x = 0 \sim 0.7$. With increasing $x$, the $T_c$ and the diamagnetic signals are strongly enhanced, and the optimal superconducting properties are obtained at $x = 0.5$. With further F doping, bulk superconductivity is suppressed. Figure 5(b) displays an enlargement of Fig. 5(a) near the superconducting transition. We defined $T_c^{mag}$ as an onset temperature and $T_c^{irr}$ as the starting temperature of bifurcation between $\chi_{ZFC}$ and $\chi_{FC}$. The $T_c^{irr}$ almost corresponds to the zero-resistivity temperature ($T_c^{zero}$) where the superconducting current appears. Both $T_c^{mag}$ and $T_c^{irr}$ show the highest at $x = 0.5$, which is consistent with the resistivity measurements.

On the basis of the obtained results, we established a phase diagram of $LaO_{1-x}F_xBiS_2$ prepared using high-pressure annealing at 600 $^o$C under 2 GPa. Figure 6 shows the established phase diagram with the determined $T_c^{onset}$, $T_c^{zero}$, $T_c^{mag}$ and $T_c^{irr}$. The optimal superconducting properties are obtained at the summit of the dome. The dome structure resembles the curvature of the $a$ lattice constant as shown in Fig. 3(a). This fact implies that the $T_c$ of $LaO_{1-x}F_xBiS_2$ correlates with the $a$ axis. In fact, the maximum $T_c$ observed in several $BiS_2$-based superconductors depends on blocking layer structure. When we focus only bulk $BiS_2$-based superconductors, namely $Bi_4O_4S_3$, $NdO_{1-x}F_xBiS_2$, $PrO_{1-x}F_xBiS_2$ and $LaO_{1-x}F_xBiS_2$, we note the tendency that higher $T_c$ appears with larger $a$ axis. [1,2,11,15] Furthermore, Xing et al. indicated that the $BiS_2$-based superconductivity is realized near the vicinity of insulating phase.[14] A larger $a$ value may enhance insulating nature and simultaneously realize higher-$T_c$ superconductivity in this family. With this respect, exploration for new $BiS_2$-based superconductors with larger blocking layers will be important. To achieve that, the high pressure technique will be a great skill.

## 4. Conclusion

We have synthesized novel BiS2-based superconductors $LaO_{1-x}F_xBiS_2$ with $x = 0 \sim 0.7$ using solid-state reaction and high-pressure post annealing. As compared to the $LaO_{1-x}F_xBiS_2$ samples prepared using only solid-state reaction, the lattice constants of the high-pressure samples were smaller. Superconducting transition was observed for $x = 0.2 \sim 0.7$, and the optimal superconducting properties were obtained for $x = 0.5$ with the $T_c^{onset}$ exceeding 10 K. The phase diagram showed an $x$-dependent superconducting dome. The evolution of dome-shaped dependence resembled the $x$ dependence of the $a$ axis. This may indicate that the correlation between the $T_c$ and the $a$ axis is essential for $BiS_2$-based superconductivity.

**Acknowledgment**

This work was partly supported by a Grant-in-Aid for Scientific Research (KAKENHI).

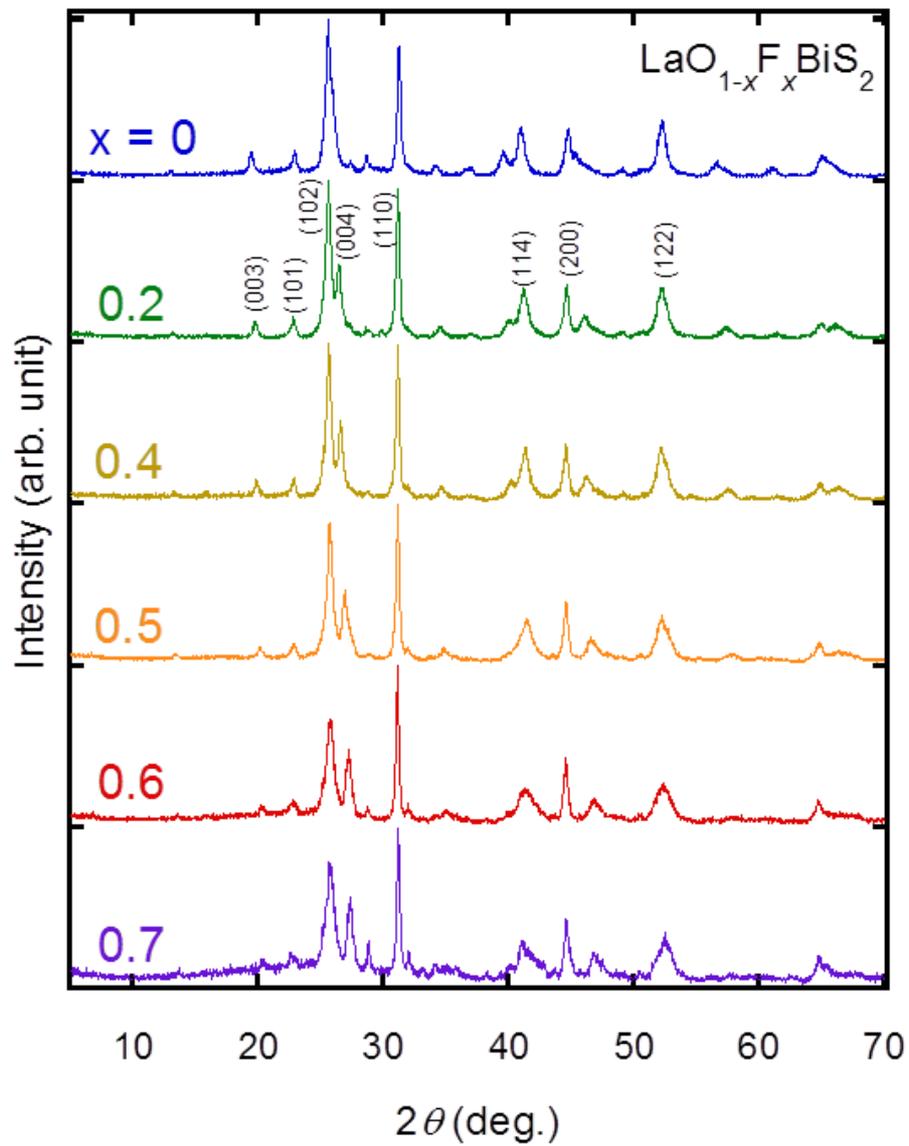

Fig. 1.   Powder x-ray diffraction patterns for LaO$_{1-x}$F$_x$BiS$_2$ with $x = 0 \sim 0.7$. The Miller indices are written in the profile of $x = 0.2$.

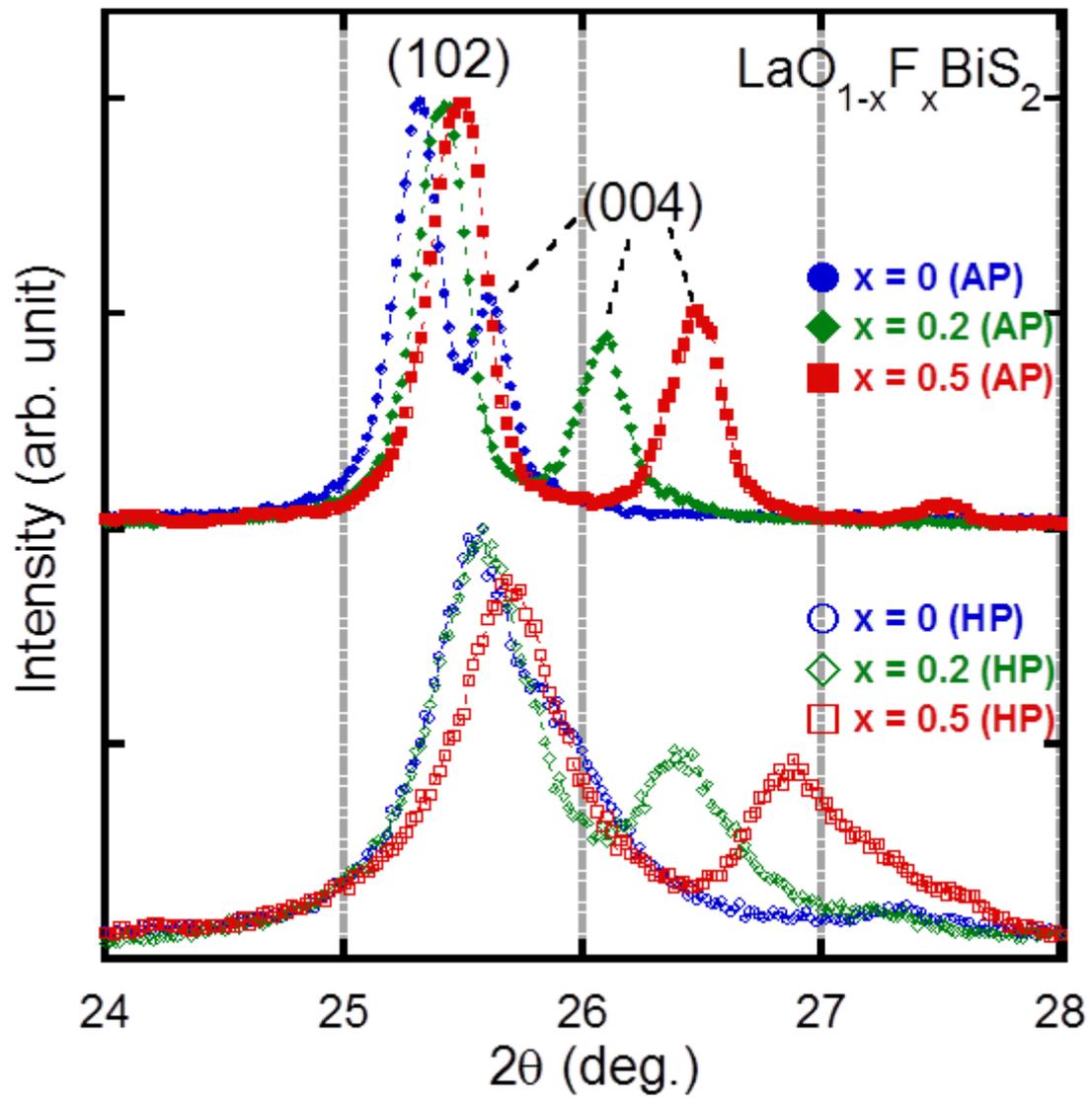

Fig. 2. Enlarged x-ray profiles near the (102) and (004) peaks for $LaO_{1-x}F_xBiS_2$ with $x$ = 0 (AP), 0.2 (AP), 0.5 (AP), 0 (HP), 0.2 (HP) and 0.5 (HP)

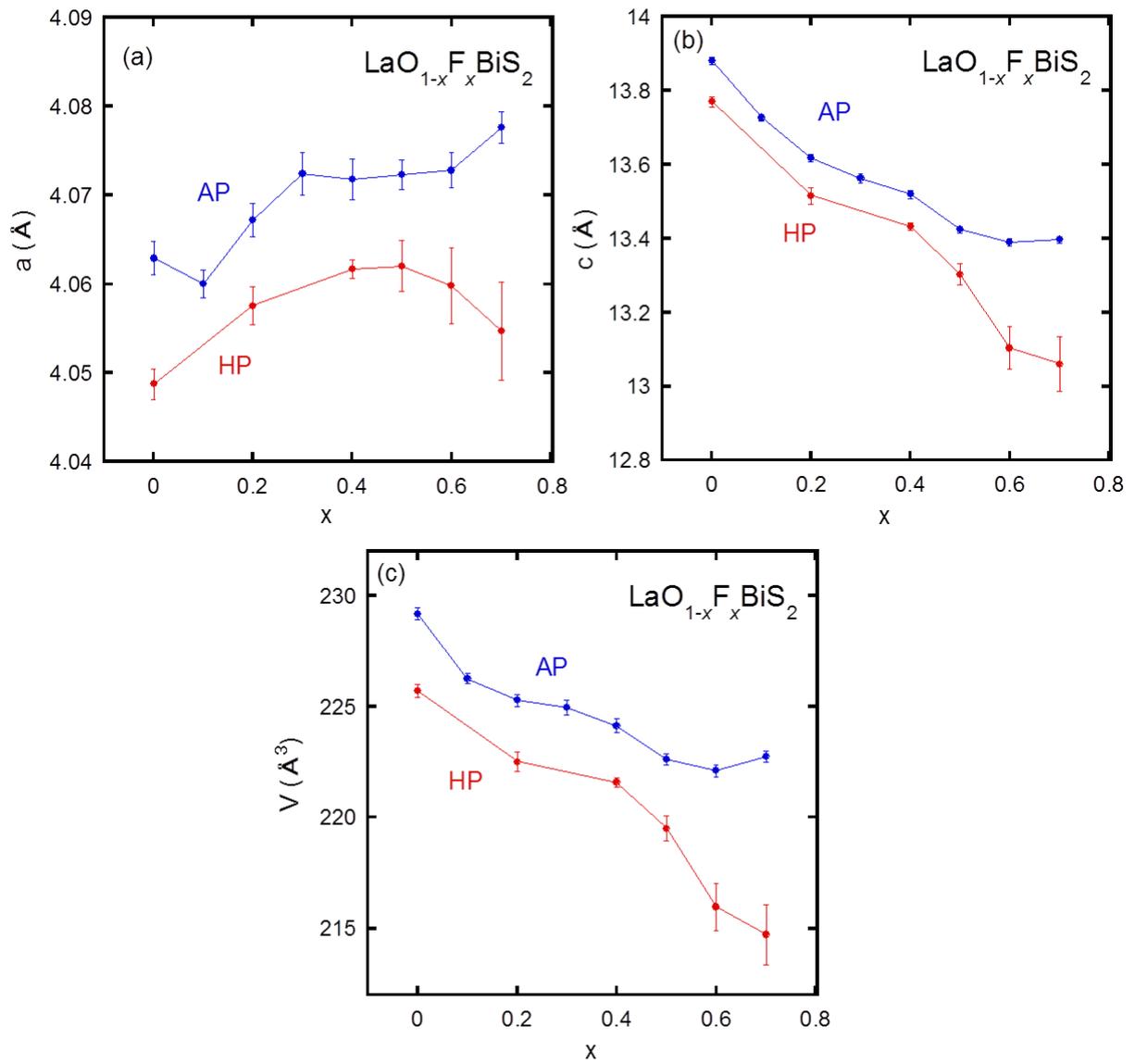

Fig. 3. F-concentration (*x*) dependence of lattice constants of (a) *a* axis, (b) *c* axis and (c) volume (*V*) for LaO$_{1-x}$F$_x$BiS$_2$ (Both AP and HP data are shown).

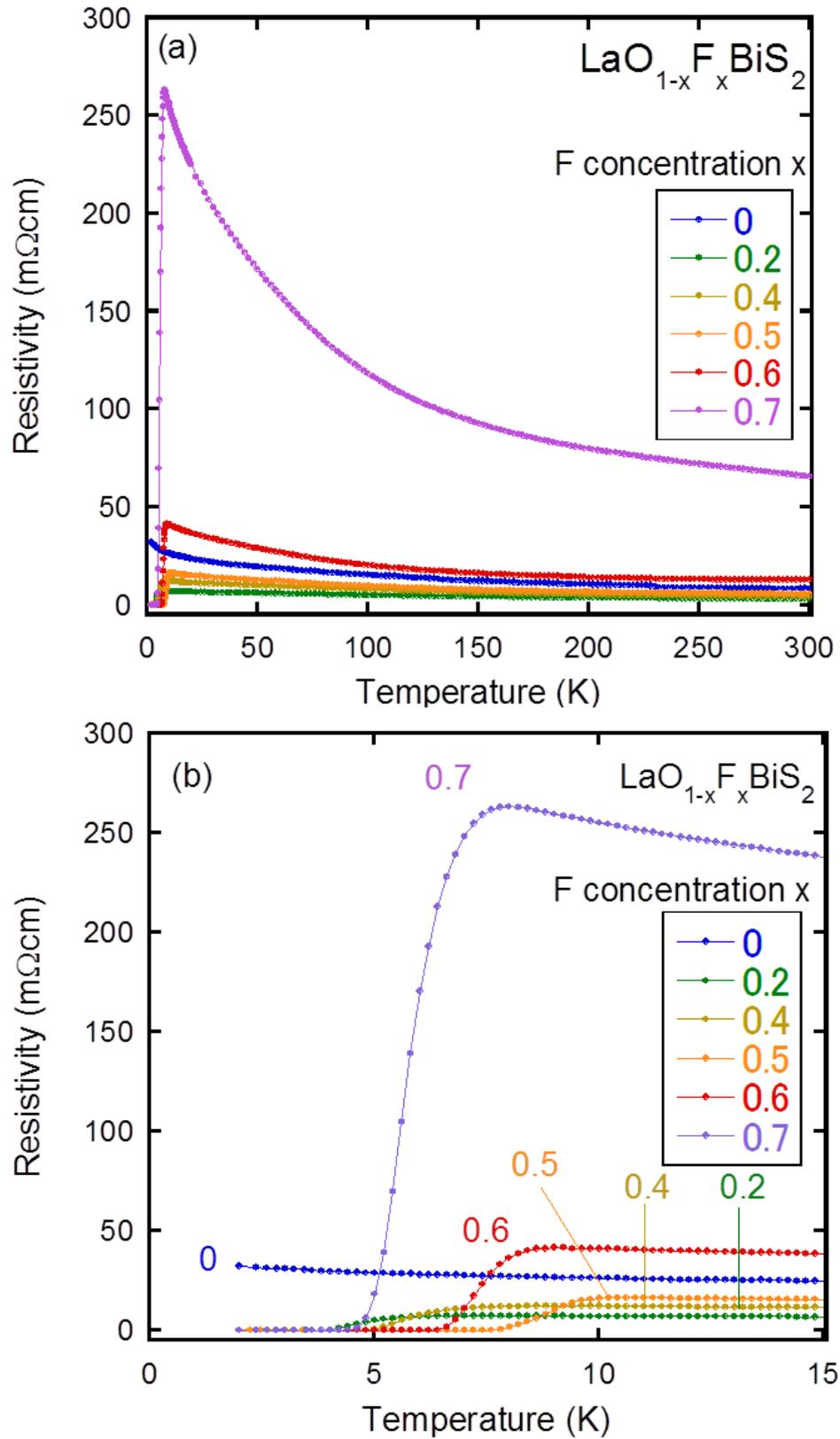

Fig. 4. (a) Temperature dependence of resistivity from 300 to 2 K for LaO$_{1-x}$F$_x$BiS$_2$ with $x = 0 \sim 0.7$. (b) Enlargement of (a) at low temperatures near the superconducting transition.

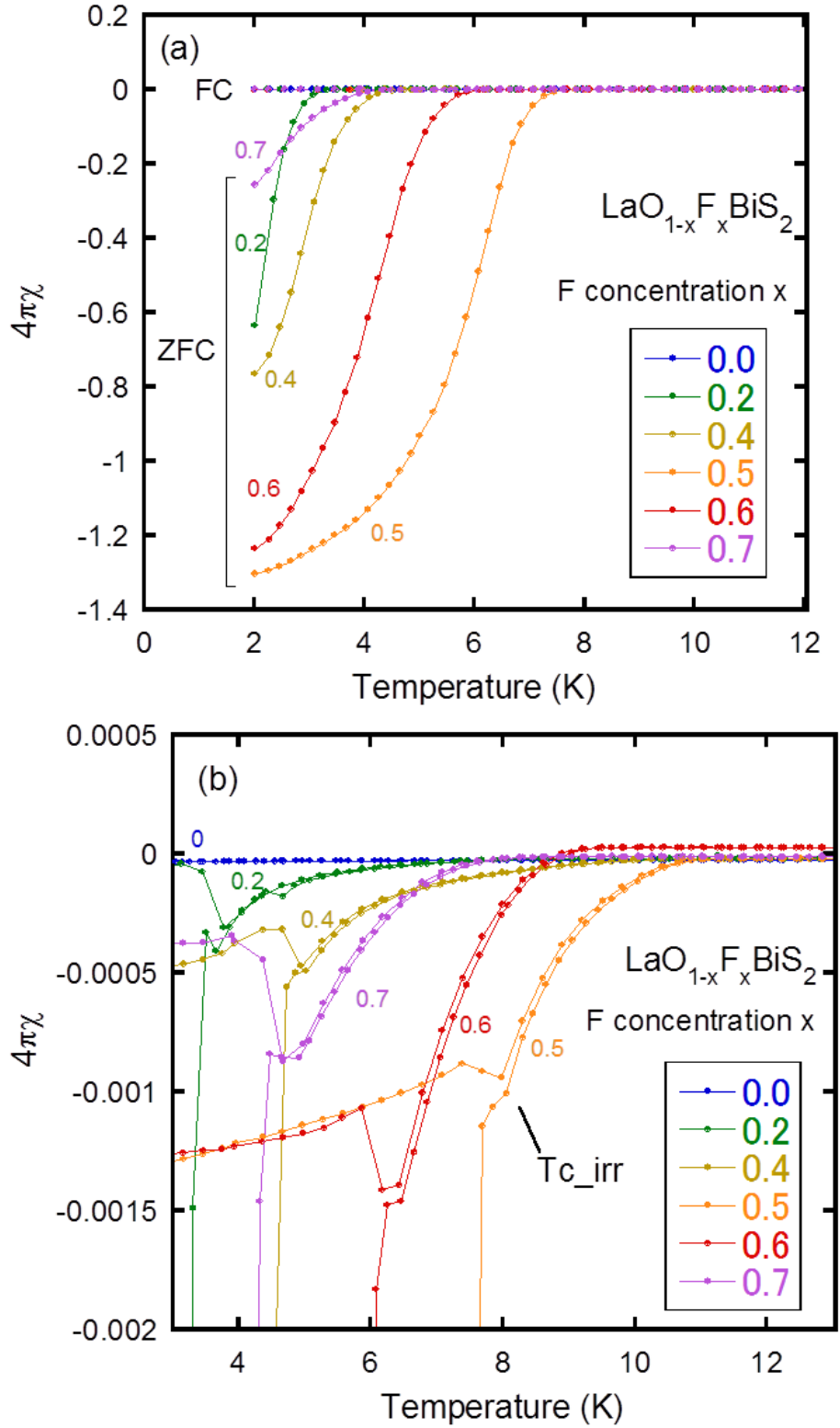

Fig. 5. (a) Temperature dependence of magnetic susceptibility from 12 to 2 K for LaO$_{1-x}$F$_x$BiS$_2$ with $x = 0 \sim 0.7$. (b) Enlargement of (a) near the onset of the superconducting transitions.

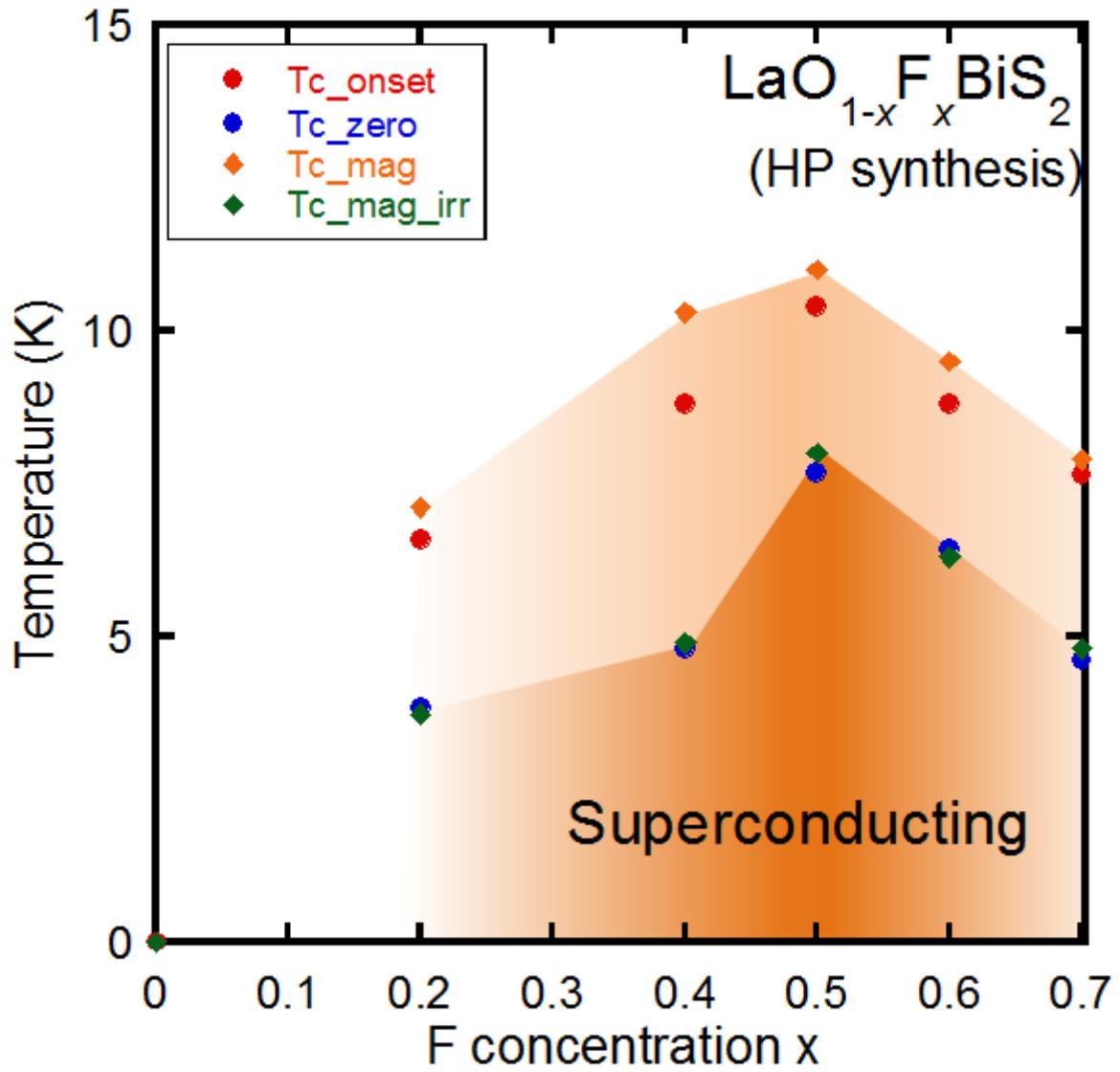

Fig. 6. Phase diagram of LaO$_{1-x}$F$_x$BiS$_2$ prepared using high-pressure annealing at 600 °C under 2 GPa.